\documentclass[twocolumn, a4paper, 11pt,book,fleqn,usenatbib]{archive}

\usepackage{graphicx}
\usepackage{txfonts}
\usepackage{color}
\usepackage[dvipsnames]{xcolor}

\usepackage{txfonts}
\usepackage{newtxtext,newtxmath}
\usepackage[T1]{fontenc}
\usepackage{graphicx}	
\usepackage{amsmath}	
\usepackage{amssymb}	

\usepackage{color}
\usepackage{multirow,bigdelim}

\begin{document} 

\twocolumn[{%
 \centering
%
{\center \bf \Large Infrared spectroscopy of clathrate hydrates for planetary science: the ethylene case}\\
\vspace*{0.25cm}

{\Large E. Dartois \inst{1}}\\
\vspace*{0.25cm}

$^1$      Institut des Sciences Mol\'eculaires d'Orsay, CNRS, Universit\'e Paris-Saclay, 
B\^at 520, Rue Andr\'e Rivi\`ere, 91405 Orsay, France\\
              \email{emmanuel.dartois@universite-paris-saclay.fr}\\

 \vspace*{0.5cm}
{keywords: molecular data, 
planets and satellites: composition, 
planets and satellites: surfaces, 
infrared: planetary systems, 
methods: laboratory: solid state, 
methods: laboratory: molecular}\\
 \vspace*{0.5cm}
{\it \large To appear in Monthly Notices of the Royal Astronomical Society}\\
 \vspace*{0.5cm}
  }]
  \section*{Abstract}
   {Hydrocarbons are observed in the gas or solid phases of solar system objects, including comets, Trans-Neptunian Objects, planets and their moons. 
In the presence of water ice in these environments, hydrocarbons-bearing clathrate hydrates could form. In clathrate hydrates, guest molecules are trapped in crystalline water cages of different sizes, a phase used in models of planetary (sub-)surfaces or icy bodies such as comets. 
The phases in presence, the potential estimate of abundances of hydrocarbon species, the spectroscopic behaviour of hydrocarbon species in the different phases must be recorded to provide reference spectra for the comparison with remote observations.  
We show in this study the specific encaged ethylene signatures, with bands similar in position, but shifted from the pure ethylene ice spectrum. They show a marked temperature dependence both in position and width. Some vibrational modes are activated in the infrared by interaction with the water ice cages.
}
%

\section{Introduction}
Ethylene is present in planets and their atmospheres,  \citep[Jupiter, Saturn, Neptune;][]{Hesman2012, Hesman2011, Romani2008, Sada2000, Varanasi2000, Maguire1997,Griffith1997, Kostiuk1993, Kostiuk1991, Kostiuk1990, Kostiuk1987}, in particular as a result of the solar photolysis of other hydrocarbons, and residing in upper atmospheres. 
Ethylene is also observed in their moons \citep[e.g., Titan, Triton;][]{Vinatier2020, Czaplinski2019, Singh2017, Beauchamp2014, Liang2010, Roe2004, Roe2001, Allen1994}.  
Hydrocarbons play an important role in particular in Titan, where methane and ethane dominate the composition and exist in the gas, and liquid/solid state, and where ethylene presence is to be expected.
C$_2$H$_4$ is present in icy bodies of the outer solar system or their atmospheres \citep[Pluto, Makemake, e.g.,][]{Brown2015, Gladstone2016}. Hydrocarbons are also observed in comets, but ethylene remains elusive in their coma.
The determination of the physical state of planetary ices and their trapping into molecular solids rely on the remote comparison to laboratory astrophysics spectra. The spectral signatures are affected by detailed solid state interactions. Water ice is an abundant component of ices, and freezes mixed with many other species. In many instances, the ice mixtures lead to simple hydrates formation with water ice in the amorphous or crystalline form. Under moderate pressure, a specific crystalline form, called a clathrate hydrate is stable, forming well defined water cages hosting the other species as guests, trapped into a crystalline network. Such a phase provides a higher stability for small species when compared to their pure phase sublimation equilibrium, providing a retention mechanism preventing their rapid and earlier escape in the presence of water ice.
The existence of ethylene clathrate hydrate in the laboratory is among the first ones reported in the literature \citep{Villard1888}. 
Small hydrocarbons like methane, ethane and ethylene clathrate hydrates crystallise under the so-called type I structure \citep{sloan2007clathrate}, i.e. in a three dimensional water ice network forming two cages, a small one with 20 water molecules with the oxygen atoms occupying the vertices of a dodecahedron (5$^{12}$), and a larger cage, with 24 water molecules with the oxygen atoms occupying the vertices of a truncated hexagonal trapezohedron possessing twelve pentagonal faces and two hexagonal faces (5$^{12}$6$^{2}$).
We explore in this article the temperature dependent spectrum of the ethylene clathrate hydrate in a range adapted to planetary science, from 6 to 160 K.

\section{Experimental methods}
\begin{figure}
\centering
\includegraphics[angle=0,width=\columnwidth]{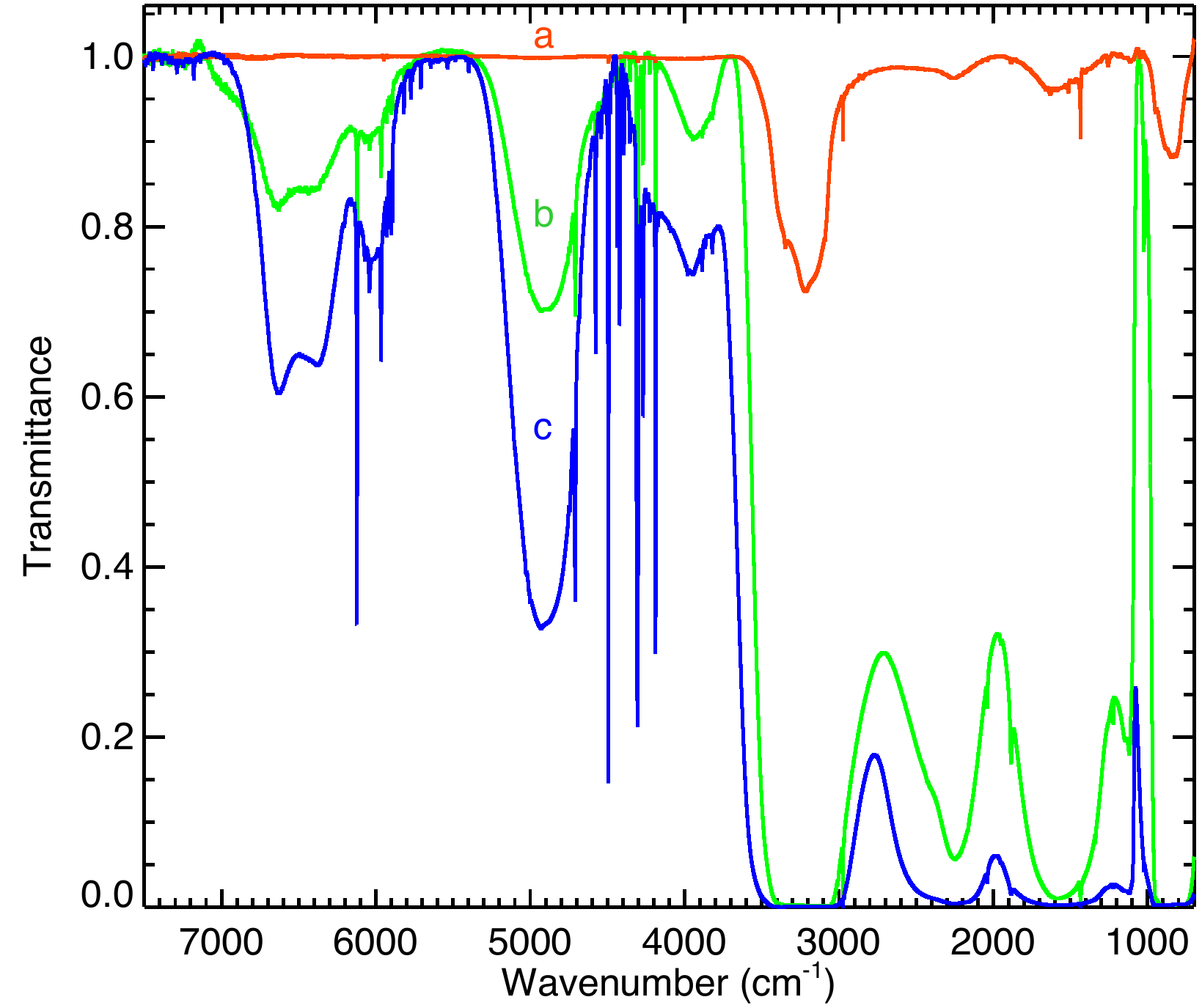}
\caption{Baseline corrected ethylene clathrate hydrate transmittance spectra of the three samples, recorded at 5.3K. Scattering effects deform the water ice bands, particularly in the second sample (b) because of imperfect film surface roughness once clathration is achieved.}
\label{Fig_generale}
\end{figure}

To form the ethylene (C$_2$H$_4$) clathrate hydrate, we use an evacuated cryogenic cell as described previously in \citep{Dartois2021, Dartois2012, Dartois2010, Dartois2009, Dartois2008} and follow a well-established protocol.
The cell is built around two ZnSe windows, facing each other, for infrared transmission analysis, sealed to an oxygen-free high thermal conductivity (OFHC), gold coated, copper closed cell. It is thermally coupled to a cold finger whose temperature can be lowered using a liquid He transfer, balanced by a surrounding Minco polyimide thermofoil heater, maintaining a constant temperature.
A high vacuum evacuated cryostat ($\rm P < 10^{-7}$mbar) surrounds the cell. Gases can be injected to the cell (or evacuated) with a stainless steel injection tube brazed at the bottom of the cell.
The clathrate hydrate formation is obtained by applying high pressure gas to an ice film. The water ice film is formed by injecting water vapour in the cell maintained just below the water ice freezing point, forming an Ih ice. 
De-ionized water was used. Several freeze-pump-thaw degassing cycles were applied to remove any dissolved gases.
The cell temperature was then lowered to 250K for the thin film and 270~K for the two thick films, to proceed for the clathrate formation. The cell was pressurised with ethylene gas at about 31 and 40 bars, for the thin and thick films, respectively, placing the system in a thermodynamic state \{pressure,temperature\} well above the stability curve for ethylene clathrate hydrate, which is about 2 bars at 250~K and 5 bars at 270~K. 
The system is kept under pressure for three to four days.
The deposited water ice film thickness is several to hundred microns thick (the film deposited on each face of the cell is about 0.1$\mu$m, 20 and 60 microns thick for the thin and two thick films considered).
An additional experiment was performed to produce a perdeuterated C$_2$D$_4$ clathrate hydrate, with 6 bars pressure and a formation at 240K, in order to elucidate potential small cage occupancies in the stretching mode region, as discussed below.
After clathrate formation, the temperature was progressively lowered (at a rate of $\sim$1K/min) while evacuating the gas.
During this operation, we maintained the system just below the sublimation/vaporisation curve of pure ethylene, and above the ethylene clathrate hydrate expected stability curve \citep{Falabella1975,Diepen1950}. 
When the system reaches a temperature sufficiently low enough so that the kinetics of the clathrate destabilisation becomes very long (typically below 100~K), the system is fully evacuated while the temperature continues to be lowered to its minimum of about 6K.
 The infrared spectra were then recorded with steps of 20K, which allows us to follow the evolution of the spectral signatures, up to the clathrate hydrate rapid dissociation occurring above 180~K over an hour under our experimental conditions. 
In between each temperature stabilised step, the temperature was raised at a rate of $\sim$1K/min, allowing for proper thermalisation of the clathrate hydrate.
Three experiments were performed with starting ice films of different thicknesses, to sample from the weakest ethylene spectral signatures in the near infrared in the 7500-3500 cm$^{-1}$ (thickest film, c) to the main vibrational modes and combinations outside the main ice bands (thick film, b) and to be able to record the ethylene signatures absorbing in the water ice OH stretching mode (thin film, a), otherwise saturated.
The baseline corrected transmittance spectra recorded at the lowest temperature are shown in Fig. \ref{Fig_generale}.
These spectra were recorded with a Bruker Fourier transform infrared spectrometer covering the 7500-700 cm$^{-1}$ spectral range at a resolution of 0.5~cm$^{-1}$, with a globar source, KBr beamsplitter and an HgCdTe detector cooled with liquid N$_2$.
%
%
%
%
%
%
%
\begin{figure*}
\centering
\includegraphics[angle=0,width=0.99\columnwidth]{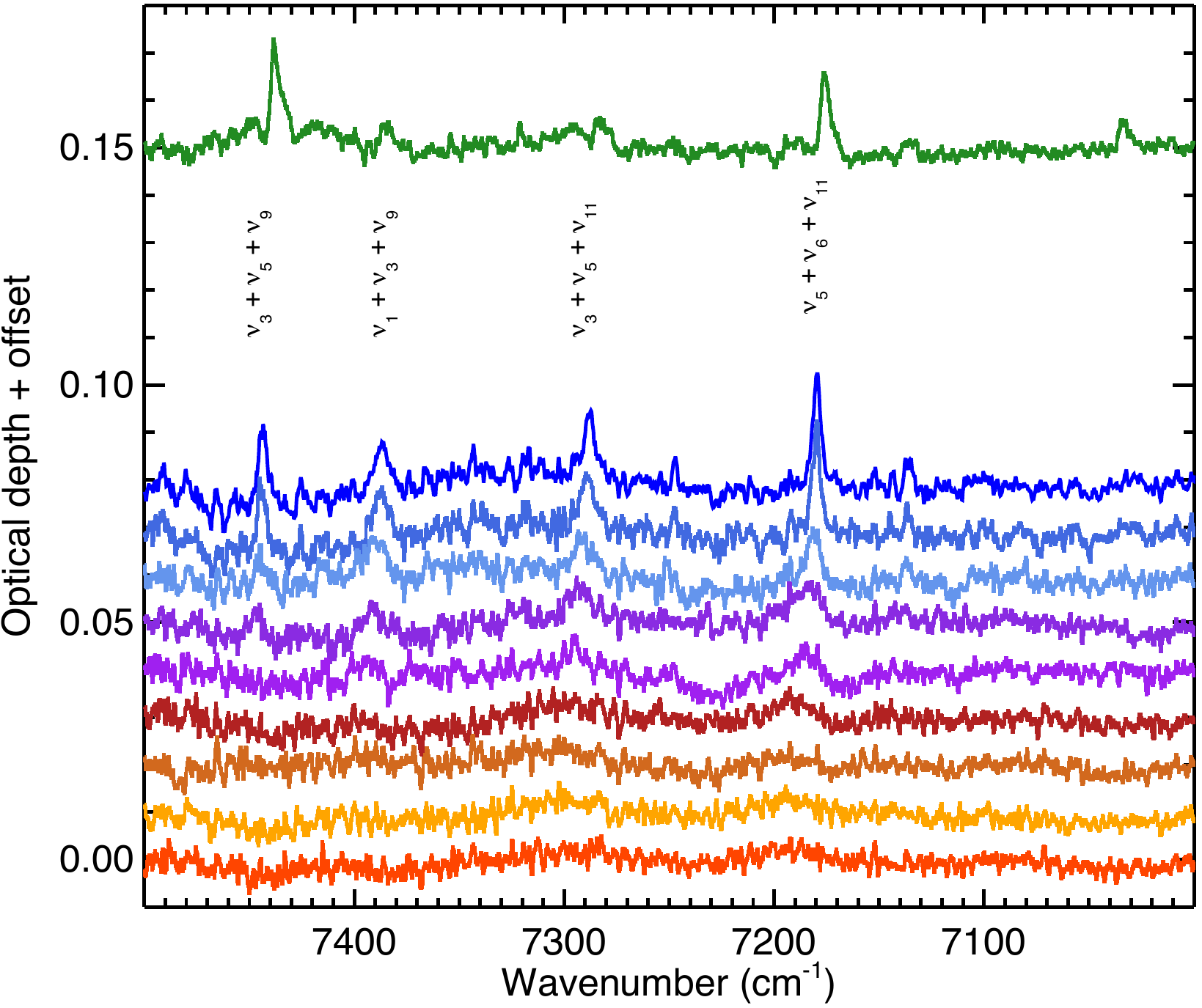}
\includegraphics[angle=0,width=0.99\columnwidth]{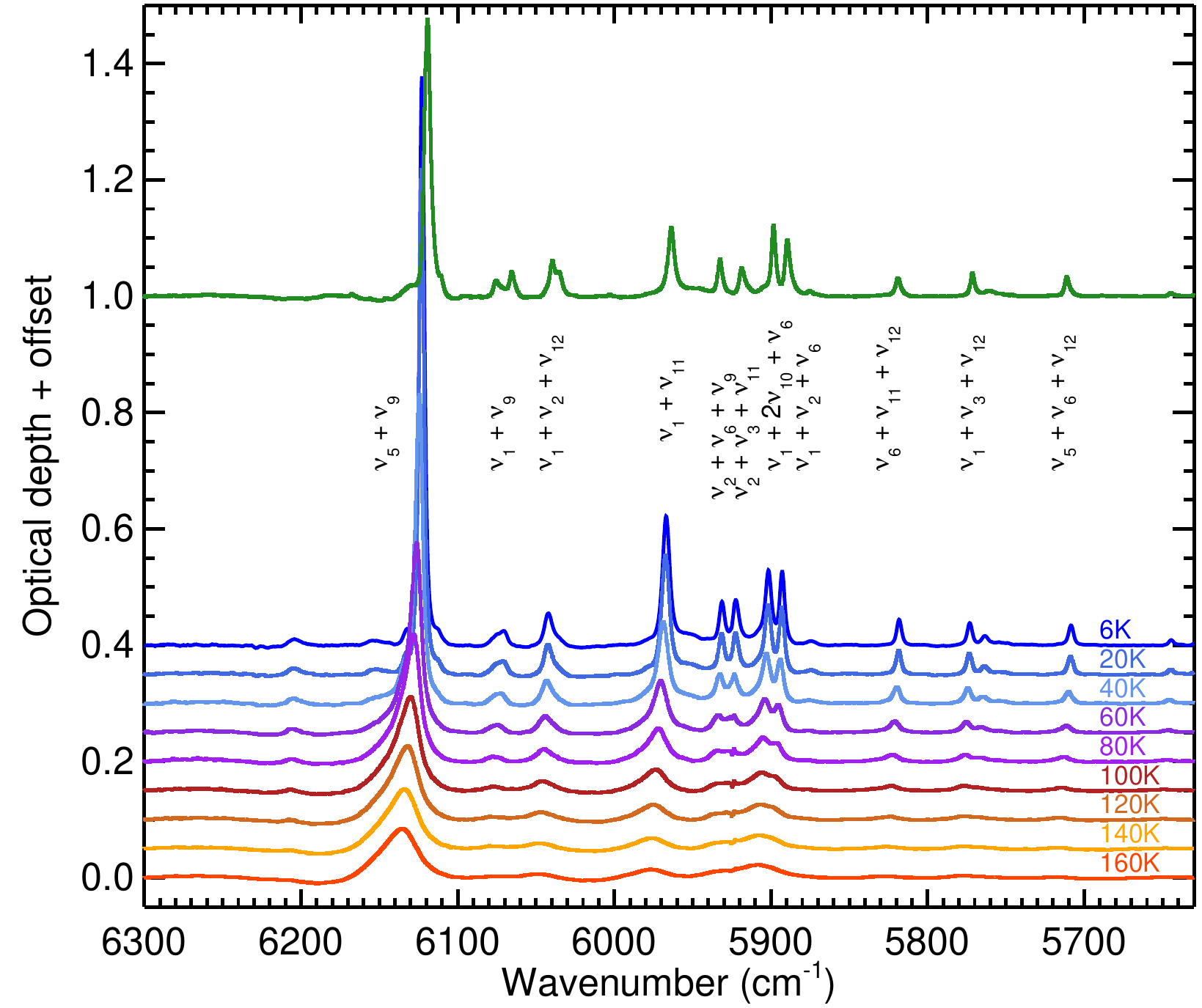}
\includegraphics[angle=0,width=0.99\columnwidth]{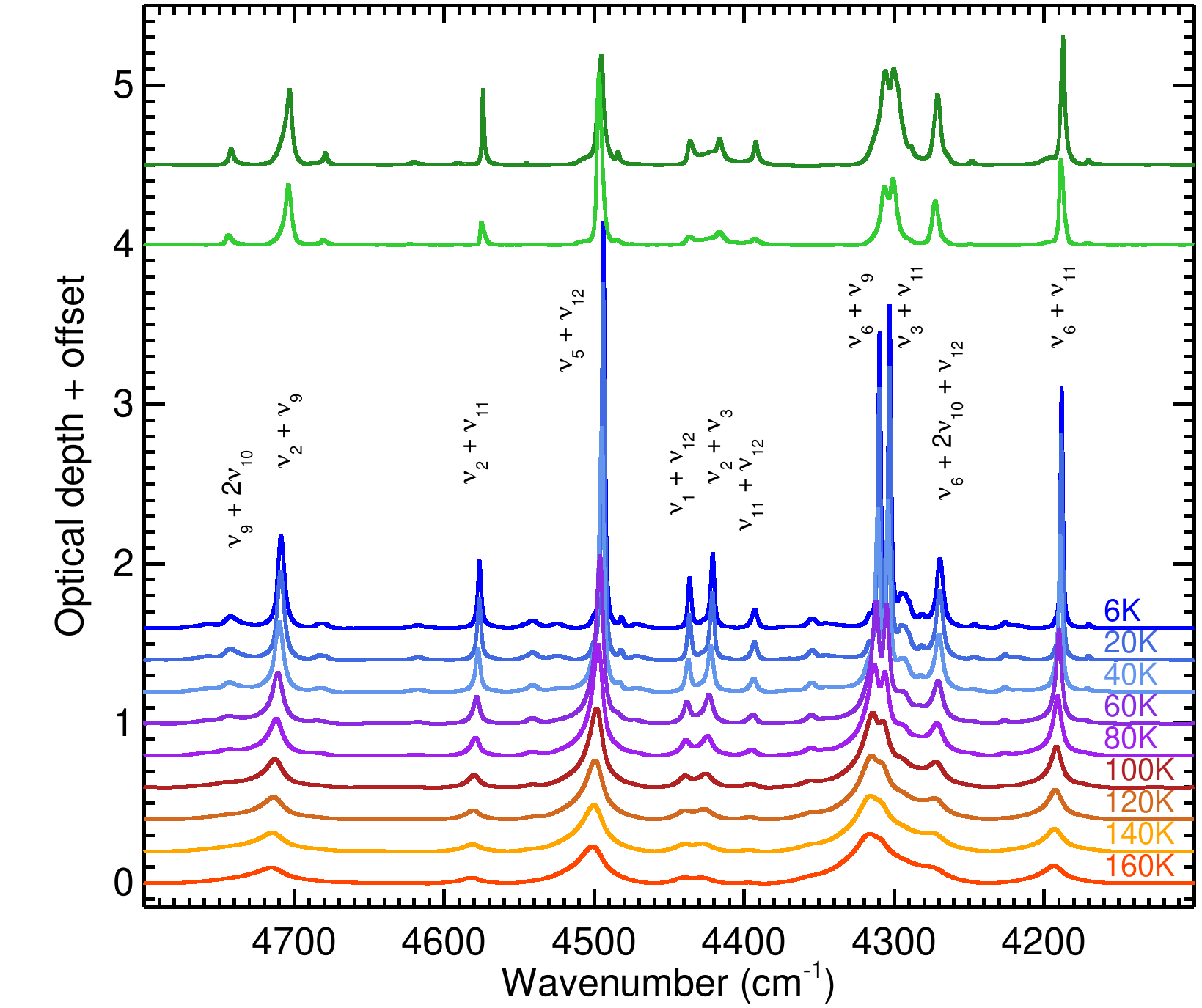}
\includegraphics[angle=0,width=0.99\columnwidth]{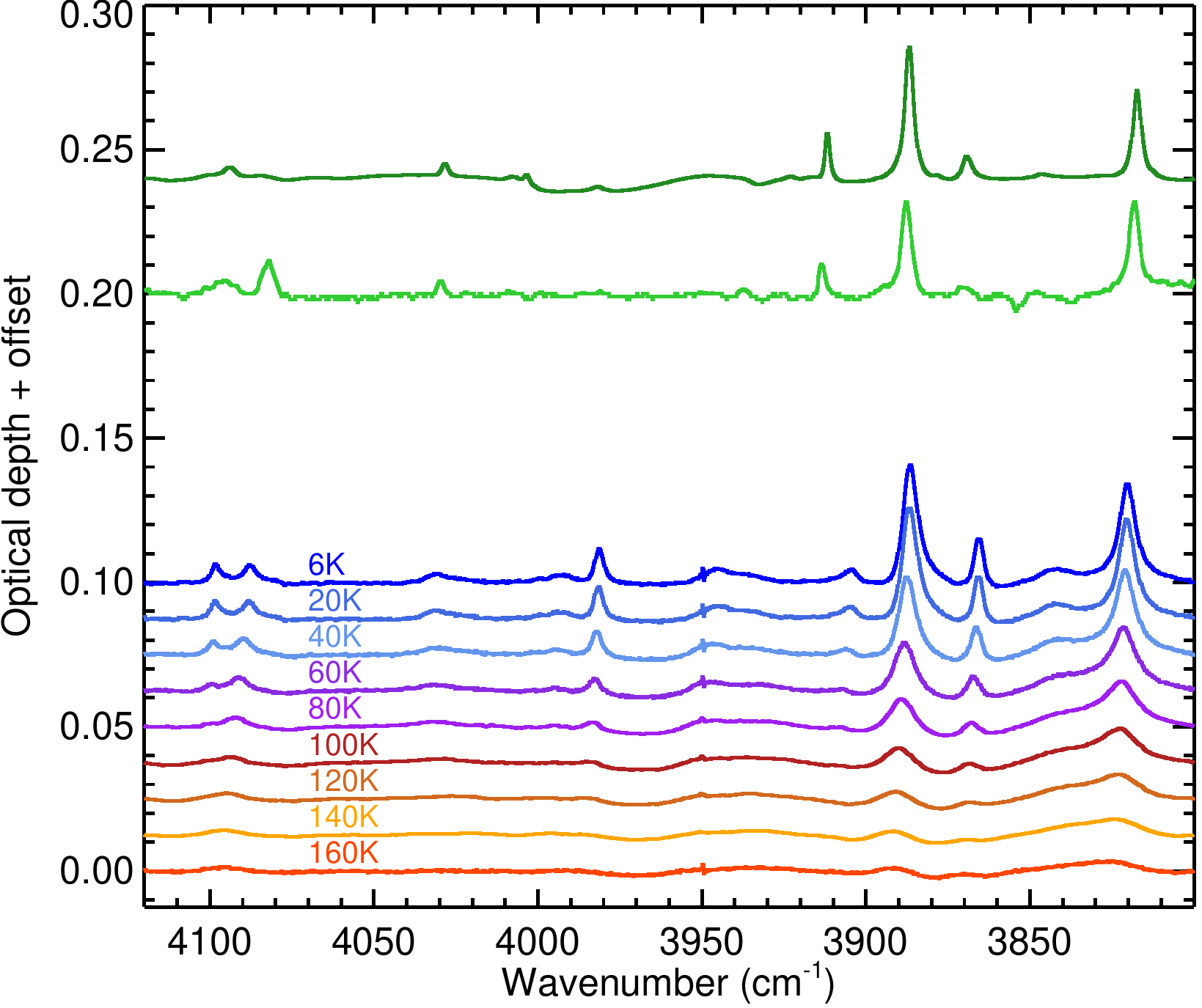}
\caption{Ethylene clathrate infrared spectra in the near infrared region involving combinations/overtones, recorded between 6K and 160K. Tentative assignments of the implied vibrational modes are given (see text for details). The crystalline pure ethylene spectrum of \citep{Hudson2014} recorded at 16K is shown just above. A pure ethylene spectrum recorded in the cell at 77K is shown on top. Spectra are offset for clarity.}
\label{Fig_NIR}
\end{figure*}
%
%
%
\begin{figure}
\centering
\includegraphics[angle=0,width=0.99\columnwidth]{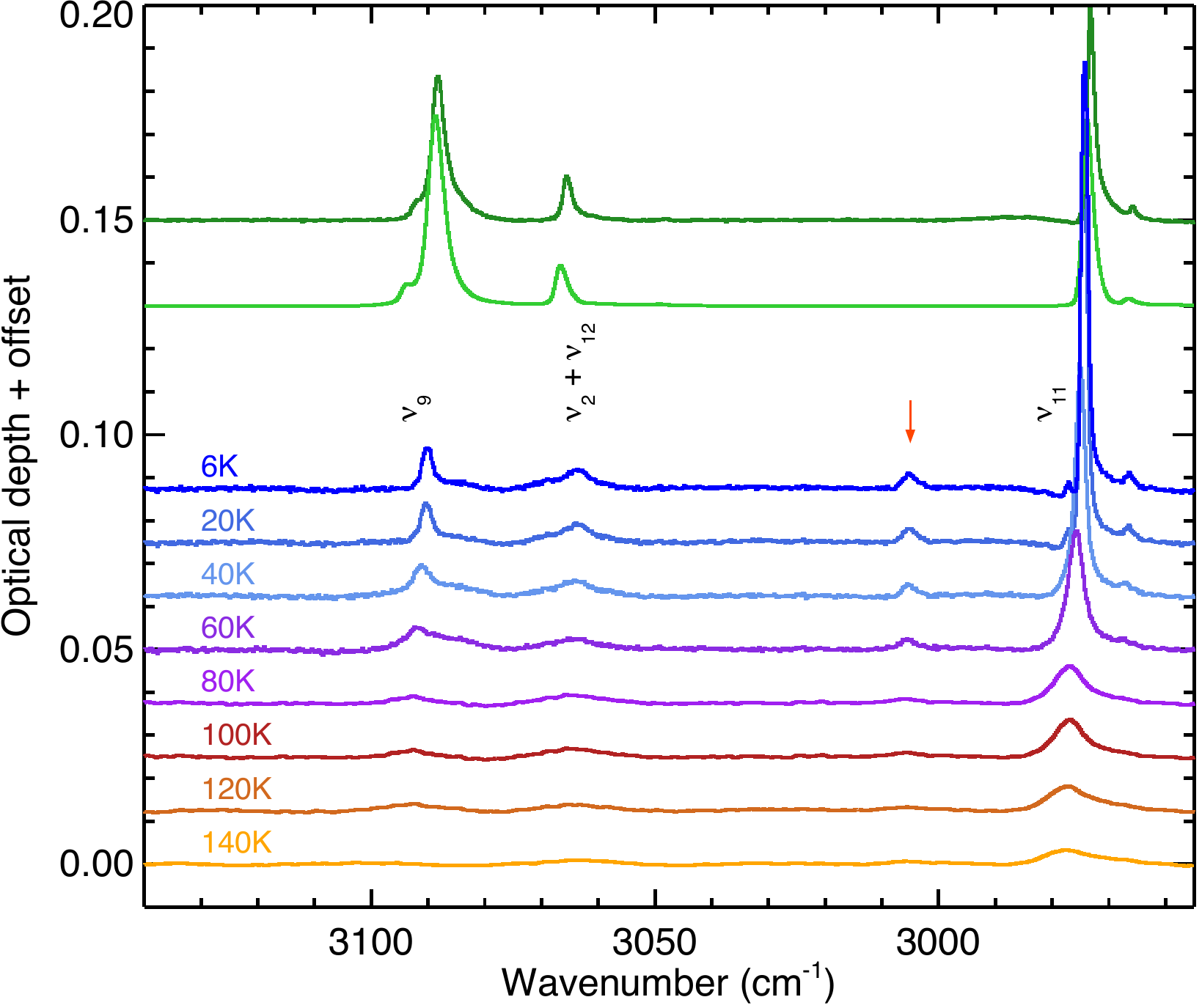}
\includegraphics[angle=0,width=0.99\columnwidth]{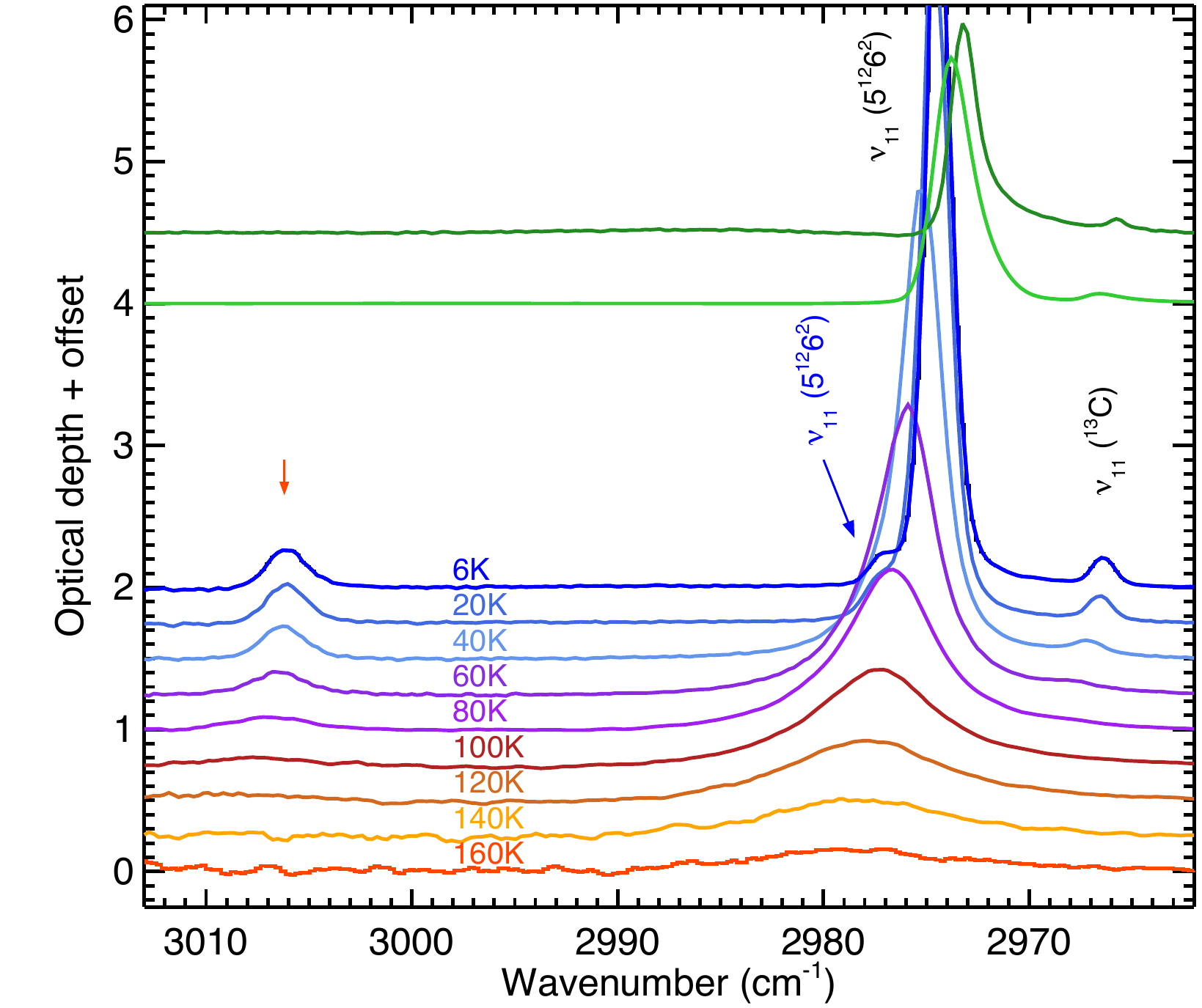}
\caption{Ethylene clathrate infrared spectra in the CH stretching mode region recorded between 6K and 160K (spectra are offset for clarity). Left panel: is the thin film allowing to access the $\nu_9$ and $\nu_2+\nu_{12}$ modes, hidden by the saturated water ice OH strong absorption in the other spectra. Right panel: thick film, with saturated C$_2$H$_4$ absorption at low temperature. Note the new band absent from pure ethylene spectra, shown with an arrow, attributed either to the $\nu_11$ mode of ethylene trapped in the small cages of the clathrate hydrate or the infrared activation of the $\nu_1$ mode. Tentative assignments of the implied vibrational modes are given (see text for details). The crystalline pure ethylene spectrum of \citep{Hudson2014} recorded at 16K is shown just above. A pure ethane spectrum recorded in the cell at 77K is shown on top.}
\label{Fig_stretch}
\end{figure}
%
%
%
%
\begin{figure}
\centering
\includegraphics[angle=0,width=0.99\columnwidth]{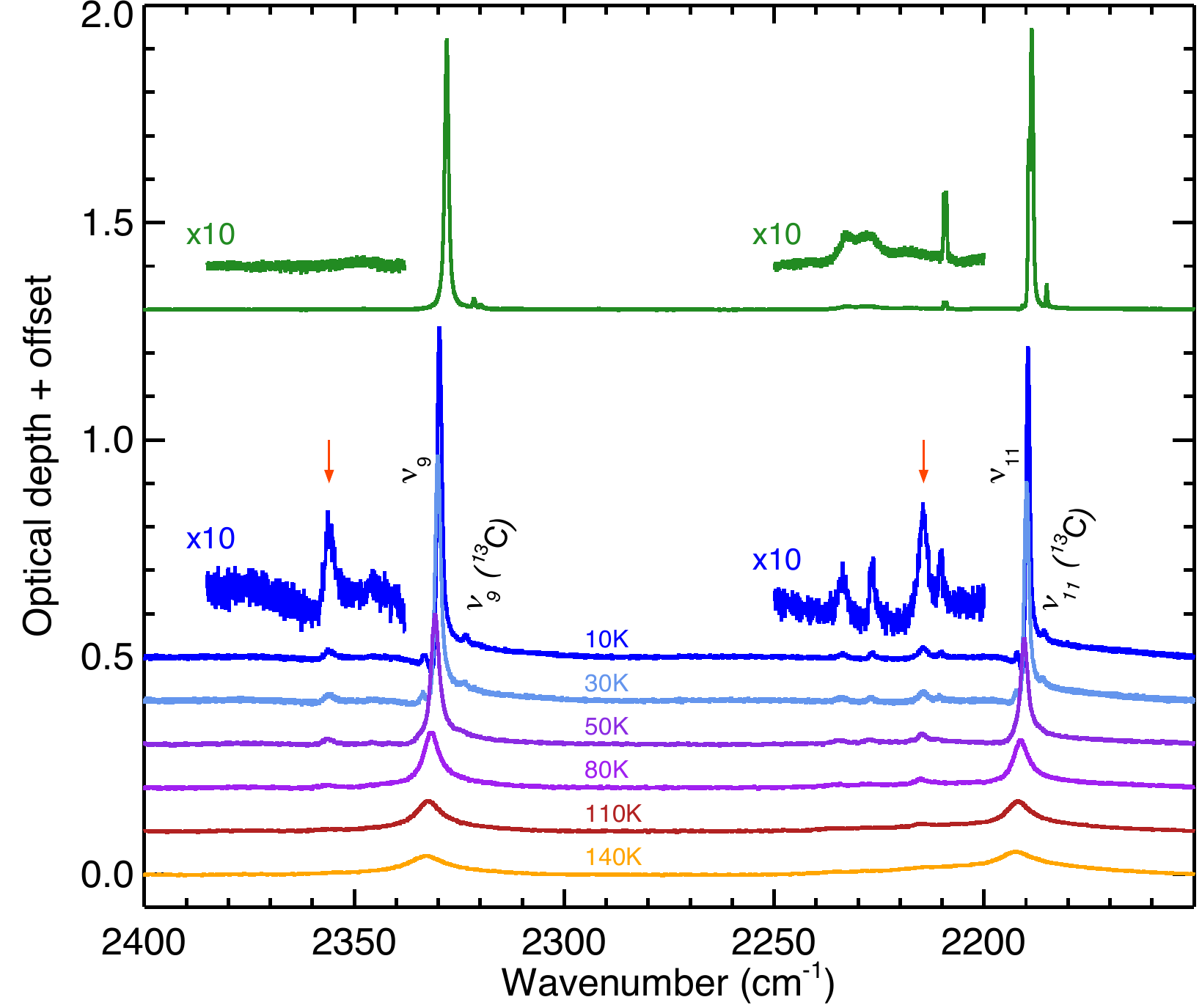}
\caption{Perdeuterated ethylene clathrate infrared spectra in the CD stretching mode region recorded between 10K and 140K (spectra are offset for clarity). The crystalline pure ethylene-D4 spectrum recorded in the cell at 10K is shown just above. Note the new bands, shown with an arrow, absent from pure ethylene-D4 spectrum. Assignments of the vibrational modes are given (see text for details).}
\label{Fig_stretch_C2D4}
\end{figure}
%
%
%
%
\begin{figure}
\centering
\includegraphics[angle=0,width=0.99\columnwidth]{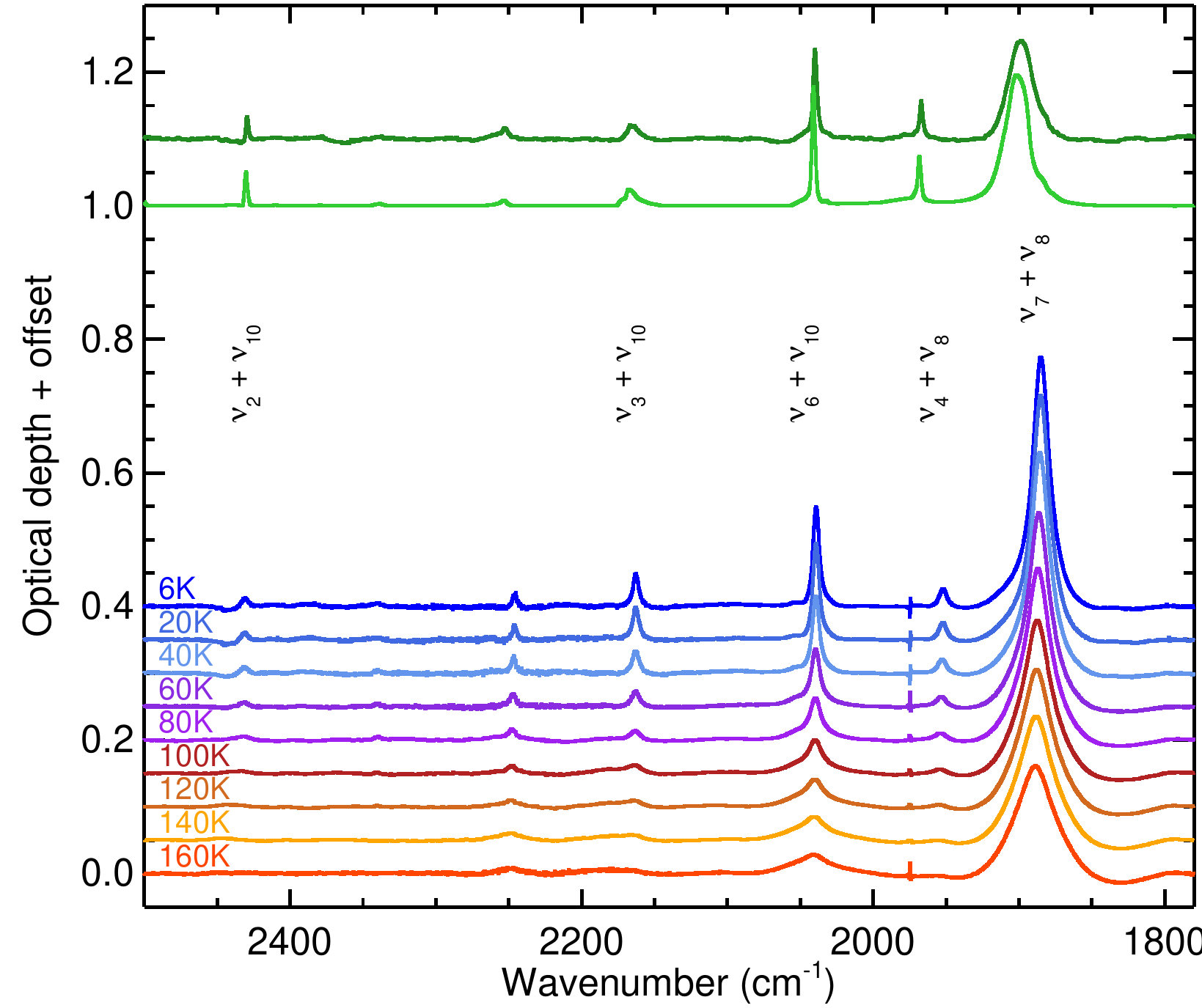}
\includegraphics[angle=0,width=0.99\columnwidth]{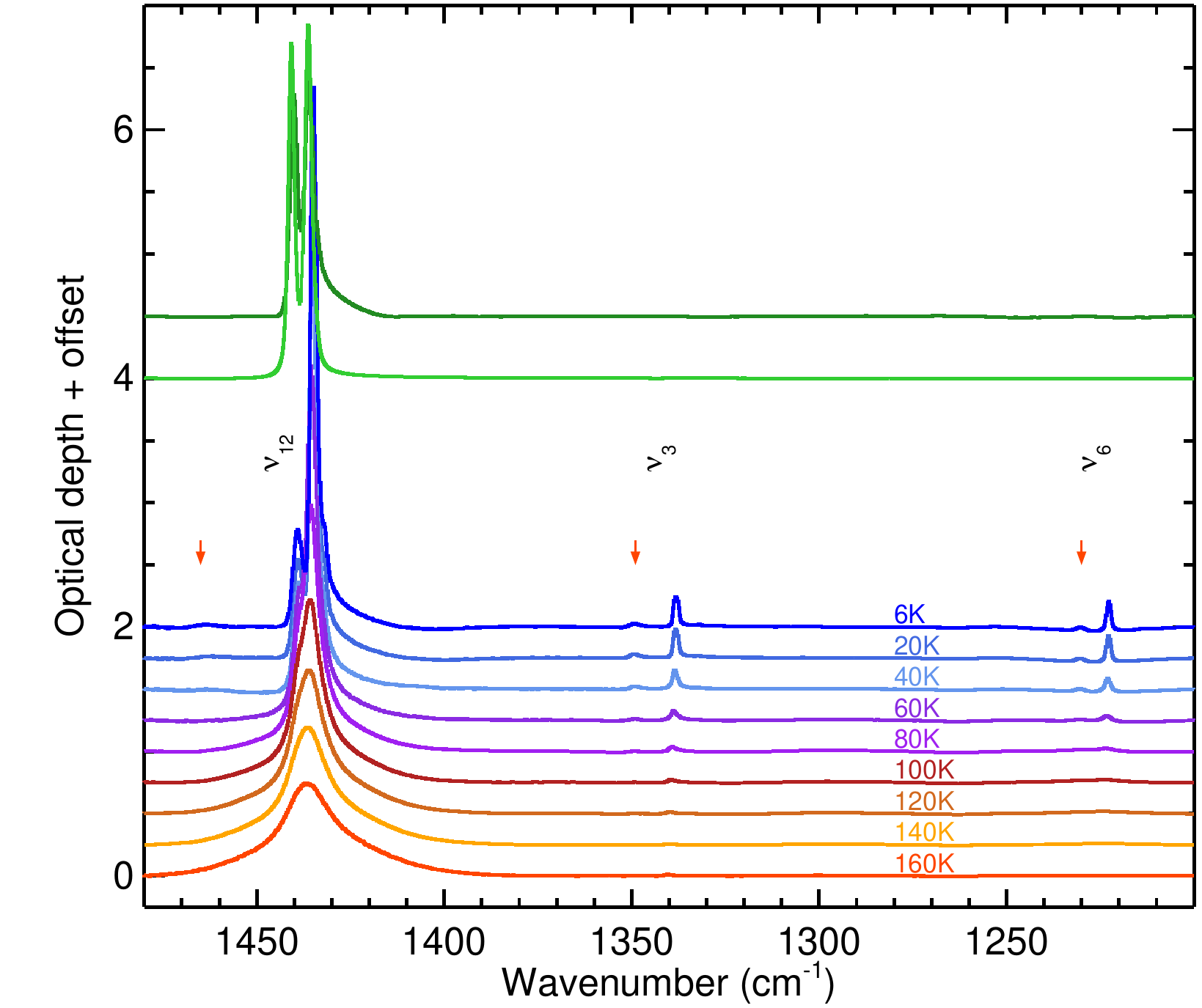}
\includegraphics[angle=0,width=0.99\columnwidth]{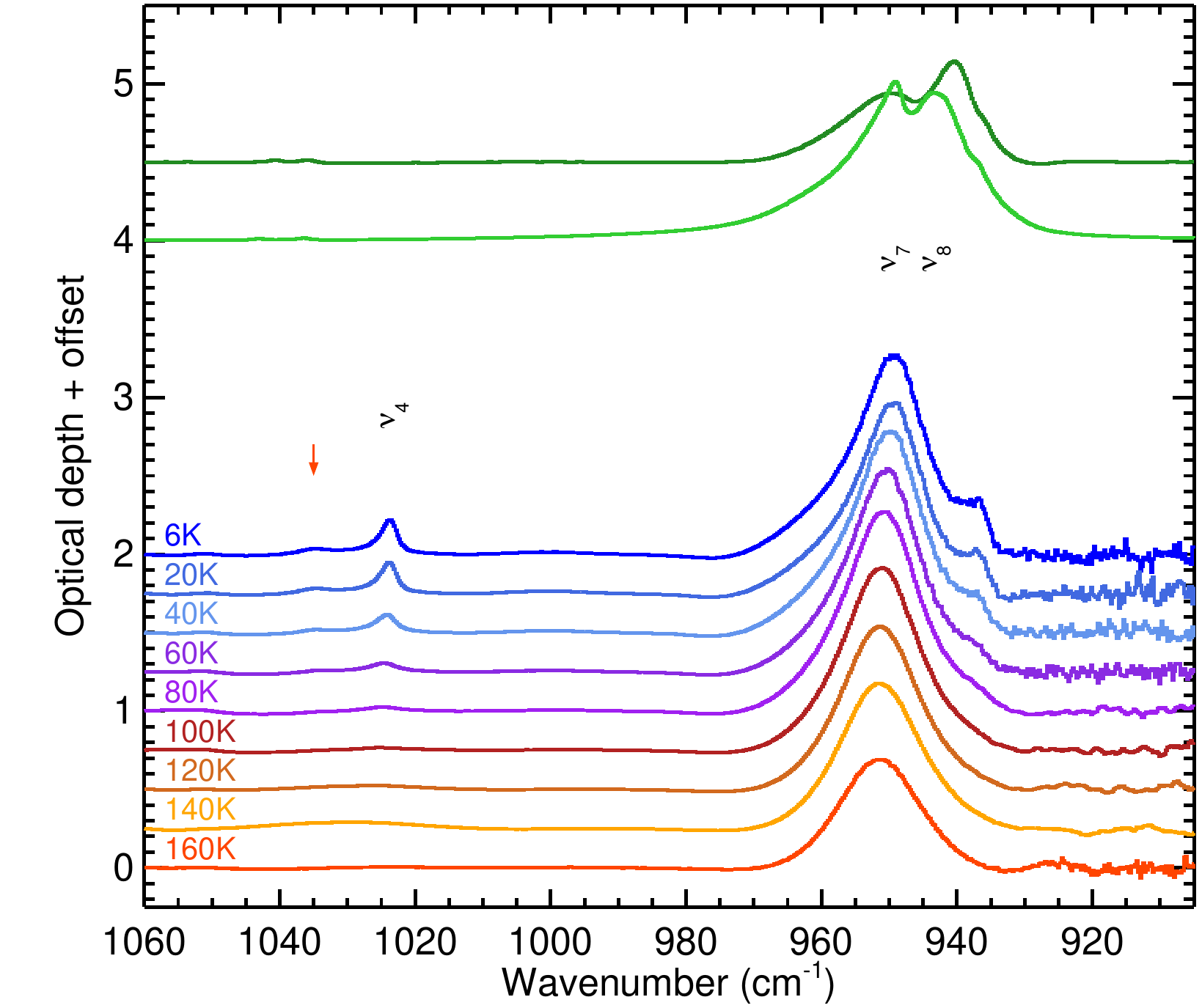}
\caption{Ethylene clathrate infrared spectra in the mid infrared region recorded between 6K and 160K (spectra are offset for clarity).  Tentative assignments of the implied vibrational modes are given (see text for details). The crystalline pure ethylene spectrum of \citep{Hudson2014} recorded at 16K is shown above. A pure ethane spectrum recorded in the cell at 77K is shown above too.}
\label{Fig_MIR}
\end{figure}
%
%
%
\section{Results}
The ethylene clathrate hydrate spectra recorded are displayed in Fig.\ref{Fig_NIR} for the near infrared combinations and overtones bands, using the thicker film to explore the weakest transitions as possible. Fig.\ref{Fig_stretch} shows spectra for two different experiments in the stretching mode region. Among these, the thinner film allows accessing the $\nu_9$ stretching mode, that, due to its position, is otherwise easily blocked in the strongly absorbing water ice OH stretching mode for thicker clathrate hydrate films. The mid-infrared bands measured are shown in Fig.\ref{Fig_MIR}. 
Above the clathrate spectra are displayed crystalline solid C$_2$H$_4$ spectra recorded in the cell either at 77K or 90K, as well as literature spectra recorded at lower temperature by \cite{Hudson2014}\footnote{available at https://science.gsfc.nasa.gov/691/cosmicice/constants.html} for comparison.
Positions and tentative vibrational mode assignments are reported in Tables~\ref{positions_NIR} and \ref{positions_MIR}. The modes where assigned by comparison with previous experiments of ethylene in solid phase or rare gas liquid phase or matrices \citep{Sunuwar2020, Abplanalp2017, Brock1994, Collins1988}. Ethylene trapped in a Kr liquid/matrix network seems to provide positions close to the observed ethylene large cage clathrate hydrate positions, facilitating such assignments.
In the combination modes, several possible assignments are often overlapping, and in crowded regions should be looked as tentative.\\
The ethylene molecule fills essentially the (5$^{12}$6$^{2}$) large cage of type I clathrate hydrate. Ethylene rather large molecular diameter makes it difficult to enter in the small (5$^{12}$) cages under normal conditions. Only under pressure or within lattice deformations, the small cage can be partly occupied.
Therefore, the overall spectrum looks to first order at low temperature close to a solid ethylene one, with the bands shifted due to the interactions with the water based cages. 
Many specific differences however appear in the spectra of the trapped ethylene molecules.\\
A small amount of ethylene molecules can occupy the clathrate hydrate small cages. In such cages, the molecule is highly stressed in a rather tight environment, and the interaction considerably blue shifts the vibrational mode.
The vast majority of ethylene molecules are trapped in the large cages, with the $\nu_{11}$ CH asymmetric stretching mode absorbing around 2174~cm$^{-1}$, but a new faint band, possibly due to the small cages (5$^{12}$) occupancy appears around 3005~cm$^{-1}$ (Fig.\ref{Fig_stretch}, shown with an orange arrow), althought the activation of the symmetric CH$_2$ $\nu_1$ stretching mode by interaction with the water cages, affecting the symmetry, is an alternative plausible possibility.
If ethylene in the small cages is responsible for this band, a simple interaction model can explain qualitatively what happens for vibrational modes with the major component perpendicular to the water cage potential, such as CH stretching modes \cite[e.g.][]{Subramanian2002, Pimentel1963}. The $\nu_{11}$ CH stretching mode would be blueshifted by about 29~cm$^{-1}$ in the small cage with respect to the large cage band position at the lowest temperature measured. A similar, although slightly smaller, shift of 27~cm$^{-1}$ has been reported using a Raman active mode by \cite{Sugahara2000} for a CH symmetric stretching mode, recorded at higher temperature (and with a higher cage occupancy ratio because of a higher pressure applied during the measurement). This measurement is also consistent with the blueshift evolution of the large cage CH stretching mode we observe with temperature (Fig.~\ref{Fig_positions}). 
To test the hypothesis of possible small cages occupancy versus fortuitous overlap with an activated mode, we performed an experiment with a C$_2$D$_4$ clathrate hydrate, and analyse the corresponding stretching mode region in the 2150-2400 cm$^{-1}$ spectral range. The results of this experiment are shown in Fig~\ref{Fig_stretch_C2D4}. We observe the occurence of new small absorption bands, blue shifted and not present in pure C$_2$D$_4$ ice. For the equivalent $\nu_{11}$ CD stretching mode, the shift is of the order of 24~cm$^{-1}$.
In the C$_2$D$_4$ clathrate hydrate case we do not expect an accidental overlap with a possibly activated $\nu_{1}$ CD stretching mode that should lie around 2250~cm$^{-1}$.
Note also that the $\nu_{9}$ and $\nu_{11}$ CD stretching modes have similar relative intensities in the ice and in the clathrate hydrate cages. This seems different in the case of the $\nu_{9}$ CH stretching mode in the C$_2$H$_4$ clathrate hydrate, which falls in the core of the OH stretching mode absorption of water ice, showing different interaction behaviour.
Other low intensity absorption bands appearing in the blue side of the most intense and isolated near infrared combination mode regions can be tentatively attributed to the small cages ethylene occupancy (e.g. around 6205, 4355, or 4225~cm$^{-1}$). However, the many possibilities of overlap with low intensity combination modes hinders a definite attribution at this stage.\\
Another difference of the trapped molecules interactions with the water cages with respect to the pure ice spectra is that some vibrational modes in the mid infrared are activated (e.g. $\nu_{3}$, $\nu_{4}$, $\nu_{6}$).
These modes are also slightly activated, although much broader, in the amorphous ice mixture of H$_2$O and C$_2$H$_4$ (20:1) recorded at 15K\footnote{Cosmic Ice Laboratory : https://science.gsfc.nasa.gov/691/cosmicice/spectra.html}.

For some of them, two bands are observed, with one much more intense and attributed to ethylene in the large cages. The large cage-small cage shift is expected to be lower for these modes involving rocking and scissoring modes. For the $\nu_{3}$ H-C-H in-plane scissoring, Raman active, a measurement by  \cite{Sugahara2000} measured a (less intense than the CH stretch) shift of 9~cm$^{-1}$. 
For this mode, weakly IR activated in the lowest temperature range in our experiments, we measure a $\sim$11~cm$^{-1}$ shift separating the strongest absorption of ethylene in large cages from what we tentatively attribute to ethylene in small cages, giving rise to a small absorption. A shift of similar amplitude (8~cm$^{-1}$) is observed for a small band appearing in the blue side of the C-C-H in-plane rocking $\nu_{6}$ activated mode.\\
 At the lowest temperatures, substructures appear in some bands. In particular a structure a few wavenumber above the main band is measured in the $\nu_{11}$ CH stretching mode region (Fig.\ref{Fig_stretch}, shown with a blue arrow) and in the equivalent $\nu_{11}$ CD stretching mode region (Fig.\ref{Fig_stretch_C2D4}). This structure disappear quickly when raising the temperature. It is reminiscent of a site dependent absorption, as observed e.g. in a Xe matrix \citep[e.g., Fig 3 in][]{Collins1988}. We tentatively attribute it to a symmetry breaking in different sites in the 5$^{12}$6$^{2}$ cage with two hexagonal and five pentagonal faces.
The $\nu_{12}$ H-C-H in-plane scissoring mode shows a pronounced substructure at low temperature.\\
A generic aspect of clathrate hydrate spectra is the absence of sudden change/phase transitions in the 6K to 160K range, as would be betrayed by the infrared signatures. The spectra evolve progressively from sharp cold solid phase trapped system signatures to broad bands as the molecule gets more mobility in the cages.
Band positions and bandwidth are temperature dependent, as the mobility of the trapped ethylene molecule increases, giving access to a temperature probe.
The position-temperature diagram for selected spectrally isolated vibrational fundamentals and combination modes in the 6K to 160K range, extracted from the measured spectra, are shown in Fig.~\ref{Fig_positions}.\\
\section{Discussion}
The photochemical processing of methane and ethane in planetary atmospheres \citep[e.g.][]{Gladstone2016} and/or ices \citep[e.g.][]{Abplanalp2016} will produce acetylene and ethylene.
Although challenging, the detection of unsaturated C$_2$H$_x$ molecules and their relative abondances is important to constrain planetary models.
In environments where methane or ethane would be trapped in a clathrate hydrate, such as Titan (sub-)surface \citep[e.g.][]{Vu2020}, ethylene is expected to be trapped too. As with ethane, the presence of ethylene would lower the boundary of the stability region for clathrate hydrate formation \citep[e.g.][]{Ma2001}.
The main clathrate hydrate structure of the ethane or methane clathrate should be preserved (type I), as ethylene will most probably not be a dominant species.
If ethylene would be a significant fraction of the guest molecules, a type I structure is still expected based on recent \citep{Sugahara2003} study showing that the methane + ethylene mixed gas hydrate constructs the structure-I unit lattice in the whole relative composition range.
Ethylene and ethane clathrate hydrates phase diagrams show close-by stability curves. They should both be present if a clathrate hydrate forms.
At the surface temperatures of cold objects like Pluto ($\sim$30-50K), Titan($\sim$90K), the ethylene bands with the higher contrast in the near infrared to be searched for, and not mixed with ethane transitions are the $\nu_5+\nu_9$ ($\sim$6130~cm$^{-1}$), $\nu_5+\nu_12$ ($\sim$4500~cm$^{-1}$), and $\nu_6+\nu_11$ ($\sim$4190~cm$^{-1}$) modes.\\
These modes will behave differently in a clathrate hydrate with respect to the signatures expected either in the pure ice \citep[this work]{Molpeceres2017, Hudson2014}, or when, e.g., diluted in nitrogen ice \citep{Quirico2007}, both because of the band shifts, but also in the information contained in the bandwidth of the considered transitions. These differences can thus provide a firm signature for the presence of a clathrate hydrate.

\section{Conclusions}
%
The ethylene clathrate hydrate near to mid-infrared spectra were recorded from 6K to 160K, a range of temperature well adapted to icy bodies conditions in the solar system, covering the 7500-700 cm$^{-1}$ ($\sim$1.33 to 14~$\mu$m) spectral range. In the absence of significant overpressure, ethylene molecules fill essentially the large cages in a clathrate hydrate ice network of type I (with small dodecahedron and large truncated hexagonal trapezohedron water ice cages) although a small fraction enters into the smaller cages. This fraction of occupied small cages will increase, in the interior of planets and their moons, if ethylene enters into clathrate hydrates formed at higher pressure.
The ethylene clathrate hydrate specific vibrational modes, and their overtones/combinations positions are tabulated, along with tentative assignments.
They can be used to be discriminated from spectra of other possible phases in presence in planetary surfaces (e.g. mixed with amorphous ice, hydrate, pure crystal, metastable phase, trapped in nitrogen matrix).
These measurements provide reference spectra to complete previous sets of experiments on hydrocarbons clathrate hydrates signatures, dedicated to methane and ethane, providing the spectral signatures for their eventual remote detection on planetary surfaces.

\section*{Acknowledgements}

This work was supported by the the Programme National "Physique et Chimie du Milieu Interstellaire" (PCMI) of CNRS/INSU with INC/INP co-funded by CEA and CNES, the CNRS/INSU and CNRS/INP. The author acknowledges the anonymous referee, as well as J\'er\^ome Guigand, Jean-Philippe Dugal and Hugo Bauduin for their support in the mechanical design.






\bibliographystyle{mnras}
\bibliography{arxiv_c2h4} 




\appendix

\section{Ethylene clathrate hydrate positions}


%
\begin{table*}
\caption{Observed ethylene clathrate hydrate positions, solid and gas phase ethylene positions and tentative assignments in the near infrared}
\begin{center}
\begin{tabular}{lllll lllll}
\hline
\multicolumn{2}{c}{$\rm \bar{\nu}_{CLH}$ (cm$^{-1}$)$^a$}  	& Tentative 	&$\rm \bar{\nu}_{Solid}^{C_2H_4}$ (cm$^{-1}$) 	&$\rm \bar{\nu}_{gas}^{C_2H_4}$ (cm$^{-1}$)$^b$\\
\multicolumn{2}{c}{}  	& assignments$^b$	&	&\\
\hline
160~K				&6~K	&		&90K(16K)	&	\\
\hline
					&7443.48	$\pm$0.81	&$\nu_{3}+\nu_{5}+\nu_{9}$	&7436.87$\pm$0.53	 	&\\
					&7386.73	$\pm$0.56	&$\nu_{1}+\nu_{3}+\nu_{9}$	&7384.66$\pm$0.55 		&7417\\
					&7288.24	$\pm$0.31	&$\nu_{3}+\nu_{5}+\nu_{11}$	&7283.10$\pm$1.23		&\\
					&7179.65	$\pm$0.10	&$\nu_{5}+\nu_{6}+\nu_{11}$	&7175.66$\pm$0.34 		&\\
					&7135.89	$\pm$0.85	&						&7136.13$\pm$1.14 		&	\\
					&6204.19	$\pm$0.50	&$\nu_{5}+\nu_{9}$ (S) ?		& 					&	\\
					&6154.04	$\pm$0.24	& 						&					&	\\
					&6132.35	$\pm$0.10	& 						&					&	\\
6136.69 $\pm$ 0.78		&6122.71	$\pm$0.10	&$\nu_{5}+\nu_{9}$			&6119.33$\pm$0.12 		&6151\\
					&6112.30	$\pm$0.50	& 						&					&	\\
					&6074.60$\pm$0.50	(sh)	&						&6075.46$\pm$0.35 		&	\\
6078.35 $\pm$ 1.33		&6070.64	$\pm$0.35	&$\nu_{1}+\nu_{9}$			&6065.58$\pm$0.20 		&\\
6049.67 $\pm$ 2.21		&6042.19	$\pm$0.20	&$\nu_{1}+\nu_{2}+\nu_{12}$	&6039.54$\pm$0.24		&6072\\
					&					&						&6035.20$\pm$0.23 		&\\
5977.11 $\pm$ 4.26		&5966.98$\pm$0.11		&$\nu_{1}+\nu_{11}$			&5963.61$\pm$0.14		&5950\\
					&5931.30	$\pm$0.43	&$\nu_{2}+\nu_{6}+\nu_{9}$	&5932.51$\pm$0.18 		&5927\\
					&5922.39	$\pm$0.20	&$\nu_{2}+\nu_{3}+\nu_{11}$	&5918.77$\pm$0.26 		&5919.7\\
5910.33 $\pm$ 2.9		&5901.65	$\pm$0.49	&$\nu_{1}+2\nu_{10}+\nu_{6}$	&5898.28$\pm$0.20 		&\\
					&5892.98	$\pm$0.14	&$\nu_{1}+\nu_{2}+\nu_{6}$	&5889.60$\pm$0.20 		&\\
				      	&5874.17 $\pm$0.11		&						&5875.14$\pm$0.53 		&\\
5828.13 $\pm$ 1.60    	&5818.25	$\pm$0.10	&$\nu_{6}+\nu_{11}+\nu_{12}$	&5819.21$\pm$0.15 		&\\
5776.06 $\pm$ 1.05      	&5773.17	$\pm$0.12	&$\nu_{1}+\nu_{3}+\nu_{12}$	&5771.48$\pm$0.14 		&5788\\
      					&5763.53	$\pm$0.18	&						&5760.39$\pm$0.14 		&\\
5718.45 $\pm$ 1.62     	&5708.56	$\pm$0.11		&$\nu_{5}+\nu_{6}+\nu_{12}$	&5711.22$\pm$0.16 		&5735\\
      					&5644.68	$\pm$0.11		&						&5644.92$\pm$0.16 		&\\
					& 4757.10 $\pm$ 0.10 	& 						& 					&\\
					& 4742.39 $\pm$ 0.10 	&$\nu_{9}+2\nu_{10}$ 		&4742.15 $\pm$0.18 (4744.00 $\pm$0.31)	&\\
4716.36 $\pm$ 0.15		& 4708.88 $\pm$ 0.10 	&$\nu_{2}+\nu_{9}$ 			&4703.10 $\pm$0.21	(4704.00 $\pm$ 0.11)	&4730.1\\
 					& 4682.85 $\pm$ 0.46 	&$2\nu_{2}+\nu_{12}$ 						&4679.23 $\pm$ 0.78 (4680.60 $\pm$ 0.48)	&\\
 					& 4680.44 $\pm$ 0.46 	& 						& 					&\\
 					& 4617.04 $\pm$ 0.16 	& 						&4620.17 $\pm$ 0.30	&\\
 4582.09 $\pm$ 0.19		& 4576.54 $\pm$ 0.21 	&$\nu_{2}+\nu_{11}$ 		&4574.13 $\pm$ 0.10 (4575.20 $\pm$ 0.15)	&4598\\
 					& 4540.87 $\pm$ 0.13 	& 						&4545.20 $\pm$ 0.10	&\\
 					& 4525.44 $\pm$ 0.10 	& 						&4526.40 $\pm$ 0.51 	&\\
 4502.05 $\pm$ 0.16		& 4493.86 $\pm$ 0.10 	&$\nu_{5}+\nu_{12}$ 		&4495.30  $\pm$ 0.15 (4496.80 $\pm$ 0.16)	&4514.8\\
 					& 4481.81 $\pm$ 0.55 	&$\nu_{5}+\nu_{12}$ ($^{13}$C) &4484.22$\pm$ 0.29 (4485.40 $\pm$ 0.54)	&\\
 					& 4471.92 $\pm$ 0.20 	& 						&4471.46	$\pm$ 0.50			&\\
 					& 4436.24 $\pm$ 0.12 	&$\nu_{1}+\nu_{12}$ 		&4436.00 $\pm$ 0.24 (4436.40 $\pm$ 0.23)	&4461.2\\
 					& 4420.82 $\pm$ 0.15 	&$\nu_{2}+\nu_{3}$ 			&4416.48 $\pm$ 0.24 (4416.60	$\pm$ 0.23)	&\\ 									& 4410.45 $\pm$ 0.28 	& 						& 					&\\
 					& 4393.09 $\pm$ 0.10 	&$\nu_{2}+\nu_{3}+\nu_{12}$/$\nu_{11}+\nu_{12}$ 		& 4392.37 $\pm$ 0.26 (4393.20 $\pm$ 0.22)	&\\
 					& 4354.52 $\pm$ 0.10 	& 						& 					&\\
 					& 4345.85 $\pm$ 0.32 	& 						& 					&\\
4316.92 $\pm$ 0.30	 	& 4309.69 $\pm$ 0.17 	&$\nu_{6}+\nu_{9}$ 			&4305.83 $\pm$ 0.65 (4306.20 $\pm$ 0.23)	&4329\\
 					& 4302.94 $\pm$ 0.17 	&$\nu_{3}+\nu_{11}$ 		&4300.04 	$\pm$ 0.57 (4300.60 $\pm$ 0.18)	&4322\\
 					& 4295.22 $\pm$ 0.27 	&$\nu_{2}+2\nu_{3}$ 		& 					&\\
 					& 4281.48 $\pm$ 0.87 	& 						& 					&\\
4273.49 $\pm$ 0.84		& 4269.43 $\pm$ 0.13 	&$\nu_{6}+2\nu_{10}+\nu_{12}$ &4270.88$\pm$ 0.17 (4272.60 $\pm$ 0.22)	&\\
 					& 4246.29 $\pm$ 0.21 	& 						&4248.22$\pm$ 0.13	 (4249.60 $\pm$ 0.29)	&\\
 					& 4226.04 $\pm$ 0.25 	& 						& 					&\\
 					& 4219.77 $\pm$ 0.21 	& 						& 					&\\
4193.74 $\pm$ 0.18 		& 4188.19 $\pm$ 0.10 	&$\nu_{6}+\nu_{11}$ 		&4187.23$\pm$ 0.13	 (4188.60 $\pm$ 0.12)	&4207\\
 					& 4170.11 $\pm$ 0.13 	& 						&4170.35$\pm$ 0.35 (4171.60 $\pm$ 0.16)	&\\
					& 4098.28 $\pm$ 0.54	&						&  					& \\
 					& 4087.67 $\pm$ 0.42	& 						& 					& \\
					& 4030.78 $\pm$ 0.29 	& 						& 					& \\
					& 3981.36 $\pm$ 0.30	& 						& 					& \\
					& 3904.71 $\pm$ 0.40 	& 						& 					& \\
3891.21 $\pm$ 2.73		& 3886.63 $\pm$ 0.13 	& 						& 					& \\					
3869.68 $\pm$ 1.00		& 3865.65 $\pm$ 0.21 	& 						& 					& \\
 					& 3841.55 $\pm$ 0.93	&						& 					& \\
3830.22 $\pm$ 2.48		& 3820.33 $\pm$ 0.22 	& 						& 					& \\					
\hline
\end{tabular}
\end{center}
\label{positions_NIR}
$^a$ At 6K, unless stated, positions uncertainties for the strongest bands are of the order of $\pm$0.2~cm$^{-1}$. 
$^b$ Tentative assignments are based on results from \citep{Sunuwar2020, Abplanalp2017, Brock1994, Collins1988}, and references therein. See text for details.
\end{table*}

%
%
%
\begin{table*}[htp]
\caption{Observed ethylene clathrate hydrate positions, solid and gas phase ethylene positions and tentative assignments in the mid infrared}
\begin{center}
\begin{tabular}{lllll lllll}
\hline
\multicolumn{2}{c}{$\rm \bar{\nu}_{CLH}$ $^a$}  	& Tentative 	&$\rm \bar{\nu}_{Solid}^{C_2H_4}$  	&$\rm \bar{\nu}_{gas}^{C_2H_4}$ $^b$\\
\multicolumn{2}{c}{(cm$^{-1}$)}  	& assignments$^b$	&(cm$^{-1}$)	&(cm$^{-1}$)\\
\hline
160~K				&6~K	&		&77K(16K)	&	\\
\hline
					& 3090.16 $\pm$ 0.16 	&$\nu_{9}$ 			&3088.23$\pm$ 0.10	 (3088.80$\pm$ 0.1)						&3105.5			\\
 	 				& 3063.88 $\pm$ 0.35 	&$\nu_{2}+\nu_{12}$ 	&3065.57 $\pm$ 0.10 (3066.60$\pm$ 0.1)						&				\\
	 				& 3005.07 $\pm$ 0.17	&$\nu_{11}$ (S) ? 		&													&				\\
2978.55 $\pm$ 0.73 		& 2974.33 $\pm$ 0.12	&$\nu_{11}$ 			&2973.25 $\pm$ 0.10 (2973.80$\pm$ 0.1)						&2988.7			\\
	 				& 2966.50 $\pm$ 0.10	&$\nu_{11}$ ($^{13}$C)	&2965.73	$\pm$ 0.10 (2966.60 $\pm$ 0.3)					&				\\
 					& 2431.10 $\pm$ 0.52 	&$\nu_{2}+\nu_{10}$ 	&2429.41$\pm$ 0.10 (2430.20 $\pm$0.10)		&				\\
					& 2340.70 $\pm$ 0.38 	& 					&(2338.40$\pm$0.21)						&				\\
2250.06 $\pm$ 2.15		& 2245.48 $\pm$ 0.15 	& 					&2252.71$\pm$0.25 (2253.40$\pm$0.19) 			&				\\
					& 2162.80 $\pm$ 0.13 	&$\nu_{3}+\nu_{10}$ 	&2164.59$\pm$2.00 (2167.80$\pm$0.25)	 blended	&				\\
2040.82 $\pm$ 0.66		& 2039.33 $\pm$ 0.14 	&$\nu_{6}+\nu_{10}$ 	&2039.86$\pm$0.10 	(2041.00$\pm$0.10)			&				\\
					& 1952.11 $\pm$ 0.11 	&$\nu_{4}+\nu_{8}$ 		&1967.06$\pm$0.12 (1968.40$\pm$0.10)			&				\\
1888.47 $\pm$ 1.43		& 1885.10 $\pm$ 0.24 	&$\nu_{7}+\nu_{8}$ 		&1898.60$\pm$0.2(1901.80$\pm$0.14)			&				\\
 					& 1463.00 $\pm$ 0.21	&$\nu_{12}$ (S) 		& 										&			\\
  					& 1439.13 $\pm$ 0.30 	&\rdelim\}{3}{5pt}[] \multirow{3}{*}{$\nu_{12}$}		&1440.10$\pm$0.34(1440.80$\pm$ 0.10)		&1443.5		\\
1436.72 $\pm$ 0.28  	& 1434.79 $\pm$ 0.17 	& 					&1436.24$\pm$0.50 (1436.40$\pm$0.1)			&				\\
  					& 1431.50 $\pm$ 0.17 	& 					&1433.50$\pm$0.50 							&				\\
 					& 1349.22 $\pm$ 0.21 	&$\nu_{3}$ (S) ? 		& 										&				\\
					& 1338.13 $\pm$ 0.10 	& $\nu_{3}$ 			& 										&1342			\\
 					& 1331.86 $\pm$ 0.10 	& $\nu_{3}$ ($^{13}$C) 	& 										&				\\
  					& 1230.38 $\pm$ 0.10 	& $\nu_{6}$ (S)			& 										&				\\
					& 1222.66 $\pm$ 0.10 	& $\nu_{6}$ 			& 										&123	6			\\
					& 1213.97 $\pm$ 0.10 	& $\nu_{6}$ ($^{13}$C)	& 										&				\\
  					& 1051.03 $\pm$ 0.10 	& 					&1040.66 $\pm$ 0.20 (1043.00$\pm$0.14) 		& 				\\
  					& 1034.63 $\pm$ 0.10 	& 					&1036.08 $\pm$ 0.41 (1036.40 $\pm$ 0.10)		& 				\\
			 		& 1023.79 $\pm$ 0.10 	& $\nu_{4}$ 			&										&1023 			\\		
951.23 $\pm$ 0.18  		& 949.54 $\pm$ 0.12		& \rdelim\}{2}{5pt}[] \multirow{2}{*}{$\nu_{7}$}	&949.057$\pm$ (949.200$\pm$0.2)			&949.3			\\
		 			& 					& 					&944.236$\pm$ (943.400$\pm$0.3)				&				\\
					& 937.25 $\pm$ 0.5 		& $\nu_{8}$			&935.5 (936.7$\pm$0.3)						&943 			\\
\hline
\end{tabular}
\end{center}
\label{positions_MIR}
SC stands for tentative assignment to C$_2$H$_4$ trapping in the clathrate hydrate small cage; $^{13}$C means a vibrational mode attributed to the $^{13}$CH$_2$=CH$_2$ isotopologue.
$^a$ At 6K, unless stated, positions uncertainties for the strongest bands are of the order of $\pm$0.1~cm$^{-1}$.
$^b$ Tentative assignments are based on results from \citep{Sunuwar2020, Abplanalp2017, Brock1994, Collins1988}, and references therein. See text for details.
\end{table*}
%
%

%
%
\begin{figure*}
\centering
\includegraphics[angle=0,width=2\columnwidth]{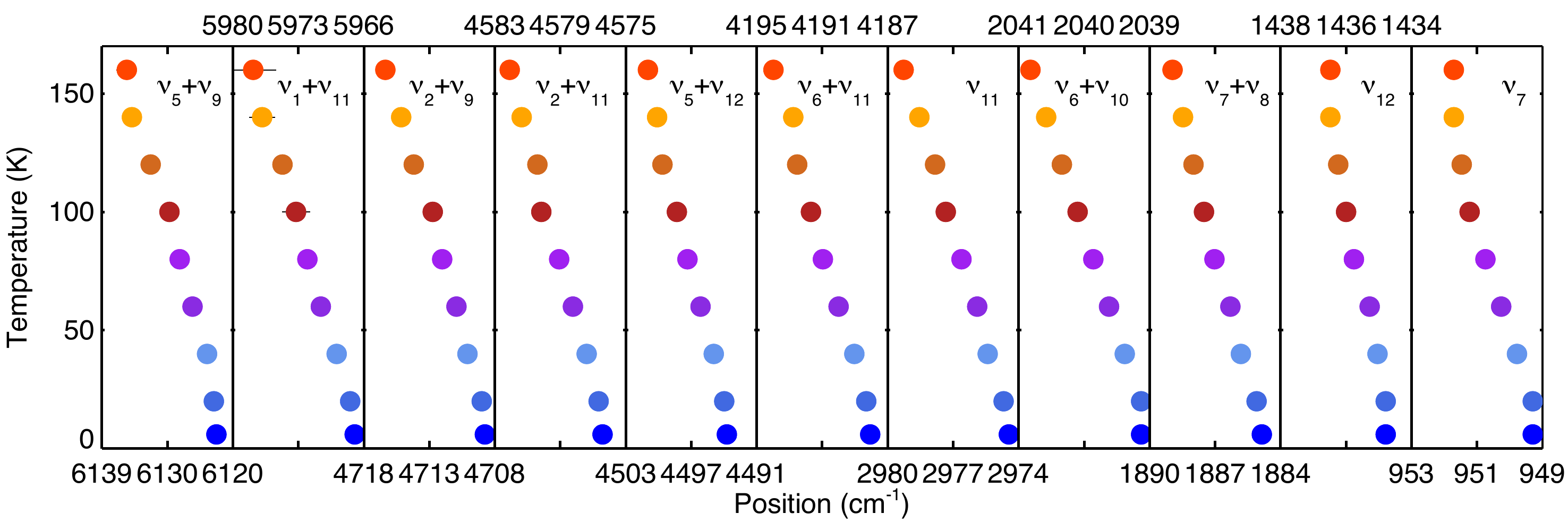}
\caption{Peak position versus temperature for selected transitions.}
\label{Fig_positions}
\end{figure*}
%
%

\end{document}